\documentclass[12pt]{iopart}
\usepackage{iopams}
\usepackage{epsfig}
\begin{document}

\title[Competition and partnership between conformity and payoff-based imitations]{Competition and partnership between conformity and payoff-based imitations in social dilemmas}

\author{Attila Szolnoki$^1$ and Xiaojie Chen$^{2}$}
\address{$^1$Institute of Technical Physics and Materials Science, Centre for Energy Research, Hungarian Academy of Sciences, P.O. Box 49, H-1525 Budapest, Hungary\\
$^2$School of Mathematical Sciences, University of Electronic Science and Technology of China, Chengdu 611731, China}
\ead{szolnoki.attila@energia.mta.hu, xiaojiechen@uestc.edu.cn}

\begin{abstract}
Learning from a partner who collects higher payoff is a frequently used working hypothesis in evolutionary game theory. One of the alternative dynamical rules is when the focal player prefers to follow the strategy choice of the majority in the local neighborhood, which is often called as conformity-driven strategy update. In this work we assume that both strategy learning methods are present and compete for space within the framework of a coevolutionary model. Our results reveal that the presence of payoff-driven strategy learning method becomes exclusive for high succer's payoff and/or high temptation values that represent a snowdrift game dilemma situation. In general, however, the competition of the mentioned strategy learning methods could be useful to enlarge the parameter space where only cooperators prevail. This success of cooperation is based on the enforced coordination of cooperator players which reveals the benefit of the latter strategy. Interestingly, the payoff-based and the conformity-based cooperator players can form an effective alliance against defectors that can also extend the parameter space of full cooperator solution in the stag-hunt game region. Our work highlights that the coevolution of strategies and individual features such as learning method can provide novel type of pattern formation mechanism that cannot be observed in a static model, hence remains hidden in traditional models.
\end{abstract}

\maketitle

\section{Introduction}
According to the evolutionary concept of game theory the fitness (payoff) of a particular strategy depends on its frequency in the population \cite{maynard_82}. During a selection mechanism this strategy becomes more or less popular depending on its success comparing to other competitor strategies. This protocol can be implemented easily via a learning process in which a player may adopt the strategy of a competitor if the latter can reach a higher payoff value \cite{sigmund_93,colman_bbs03,perc_pr17}. Naturally, this implementation allows us to extend the potential target systems from biology to more complex human populations where learning from others is an essential way to build highly cooperative societies \cite{nowak_11}.

Humans, however, are not only motivated to reach a higher payoff when making a decision who to follow during an elementary change. For example, conformity is a well-documented and frequently observed attitude among humans \cite{bernheim_jpe94,fiske_09}. In the latter case a player prefers to follow the behavior of the majority of neighboring partners, which is partly motivated by the fear to avoid too risky individual choice. Notably, conformity also plays an important role in opinion dynamics \cite{yang_hx_cpc15,yang_hx_epl16}.
Previous works already proposed the simultaneous presence of payoff-driven and conformity-driven strategy learning methods, but all of them assumed a fixed ratio of these learning protocols and explored how the collective behavior depends on this ratio \cite{pena_pre09, molleman_pone13, szolnoki_rsif15, xu_b_csf15, javarone_epl16, yang_hx_csf17}.

In the present work we assume that this ratio is flexible and we allow the mentioned learning protocols to compete for space. This extension has several practical motivations. First, it helps us to identify the specific conditions which make one of the learning protocols to be exclusive as a result of an evolutionary process. Second, such kind of coevolutionary model, where not only strategies but also learning attitudes can be exchanged between players, may offer new ways how strategies compete.
Indeed, conformity or payoff-driven learning protocol may prevail the whole system at specific parameter values. Furthermore, in some cases these strategy learning protocols are not properly competing, but they form a strategy alliance for better evolutionary outcome. In this way our present observations warrant that the diversity of strategy updating protocols could be a prime mechanism to maintain cooperation among selfish agents.

The rest of our paper is organized as follows. First we  proceed with presenting the details of our coevolutionary model that is followed by the presentation of our main results. Finally we discuss their wider implications and some potential directions for future research are also given.

\section{Coevolution of strategies and learning protocols}

For simplicity we study evolutionary social dilemma games on a square grid, but we stress that the key observations remain valid if we replace square lattice by other interaction graphs, including random networks. In the beginning each player is designated as cooperator or defector with equal probability and pairwise interactions with neighbors are assumed. Here mutual cooperation yields the reward $R$, mutual defection leads to punishment $P$, while a cooperator collects a sucker's payoff $S$ against a defector who enjoys temptation value $T$. In agreement with the widely accepted parametrization of social games we fix $R=1$ and $P=0$, while the remaining $T$ and $S$ parameters determine the character of the social dilemma \cite{szabo_pr07, perc_bs10}. More precisely, in case of $T>1$ and $S>0$ we consider a snowdrift game, but $0<T<1$ and $S<0$ result in a stag-hunt game situation. For prisoner's dilemma game $T>1$ and $S<0$ values are assumed.

Beside the mentioned strategies a player also possesses a tag, a personal feature that determines how she learns a strategy from others. In particular, we assume that a player updates strategy either in a payoff-driven or in a conformity-driven way. In the former case player $x$ acquires her payoff $\Pi_x$ by playing the particular game with all her neighbors. Furthermore, player $x$ chooses a neighboring $y$ player randomly who then also acquires her payoff $\Pi_y$ in a similar way specified above. After player $x$ adopts the strategy $s_y$ from player $y$ with a probability 
\begin{equation}
\Gamma (\Pi_x-\Pi_y) = {\{1+\exp[(\Pi_x-\Pi_y)/K]\}}^{-1}\,,
\label{adopt}
\end{equation}
where $K$ quantifies the uncertainty of strategy adoption. Without loosing generality we use $K=0.1$ that allows us to compare our results with previous findings \cite{szabo_pre05}. If player $x$ learns in a conformity-driven way then she prefers to adopt the strategy that is most common in the neighborhood of her interaction range \cite{szolnoki_rsif15}. More precisely, this player adopts the strategy $s_x$ with the probability $\Gamma (N_{s_x}-k_h) = {\{1+\exp[(N_{s_x}-k_h)/K]\}}^{-1}$, where $N_{s_x}$ is the number of players who represent $s_x$ within the interaction range of focal player $x$, whereas $k_h$ is one half of the degree of player $x$. As previously, here $K$ determines the uncertainty of learning process, which makes also possible the adoption of the strategy of the minority with a small probability. For simplicity we used the same noise value as for the pairwise imitation step. In the beginning, similarly to the strategy distribution, each player is designated as a learner using payoff-driven ($p$) or conformity-driven ($o$) motivation with equal probability. 

Technically it means that we have a four-state system where a player is payoff-driven cooperator ($p_C$), payoff-driven defector ($p_D$), conformity-driven cooperator ($o_C$), or conformity-driven defector ($o_D$). The key feature of our coevolutionary model is that players may adopt not only strategies from a neighbor, but also the way of learning. This adoption from player $y$ for player $x$ happens with the probability defined by Eq.~\ref{adopt}. In an elementary Monte Carlo step ($MCS$) a player is selected for a strategy adoption and independently for changing the learning method. These adoptions happen with the probabilities defined above. In a system containing $N$ players a full $MCS$ consists of $N$ elementary steps, hence on average all players have a chance to change strategy and/or learning method. All simulation results are obtained on interaction graphs typically comprising $N=2\cdot10^5 - 8\cdot10^6$ nodes where the stationary fractions of different states are averaged after $10^5-10^6$ $MCS$s of relaxation. The final results are averaged over 100 independent realizations for each set of parameter values.

\section{Results}

\begin{figure}
\centering
\includegraphics[width=0.7\linewidth]{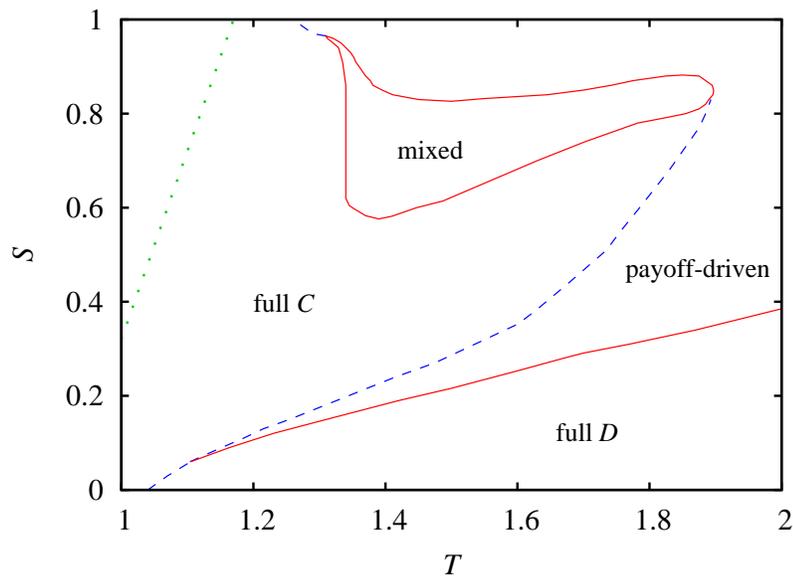}
\caption{Full $T-S$ phase diagram of snowdrift game where payoff-driven and conformity-driven strategy learning protocols are competing for space. Red solid lines denote continuous phase transitions, while blue dashed line denotes first-order phase transition. Full $C$ (full $D$) label marks the parameter regions where only cooperator (defector) strategies survive in the stationary state. ``Payoff-driven" label marks the region where payoff-driven strategy learning protocol prevails and related cooperator and defector players form a stable winning solution. The phase denoted by ``mixed" label shows where all learning protocols and all strategies coexist. For comparison, green dotted line marks the border of full cooperator state when only payoff-driven learning protocol is available for players in a uniform system.}
\label{phd_sd}
\end{figure}

We first present results obtained for the snowdrift game quadrant of $T-S$ plane where $T>1$ and $S>0$. The final outcome of the coevolutionary process for different $(T,S)$ parameter pairs is summarized in the phase diagram plotted in Fig.~\ref{phd_sd}. This diagram highlights that the region of full cooperator state is extended dramatically comparing to the basic model where players adopt external strategy only in a payoff-driven way. For comparison the border of full cooperator phase for the latter case is marked by dotted green line in the diagram. This comparison underlines that the application of a coevolutionary rule in the updating protocol can enhance the kingdom of cooperation even for snowdrift game where it was impossible to detect cooperation promoting mechanism in the framework of the traditional model \cite{hauert_n04}.

Turning back to the diagram, if $T$ and/or $S$ are too high then payoff-driven strategy learning protocol becomes exclusive and the system evolves into a state where $p_C$ and $p_D$ coexist in a role-separating way that provides optimal total payoff for the population \cite{szabo_jtb12}. Interestingly, there is an island in the parameter space where all the available states can survive and form a stable solution. This region is denoted by ``mixed" label in the diagram. As expected, if $S$ is too small and $T$ is too high then both types of cooperators die out and only defectors survive.

\begin{figure}
\centering
\includegraphics[width=0.45\linewidth]{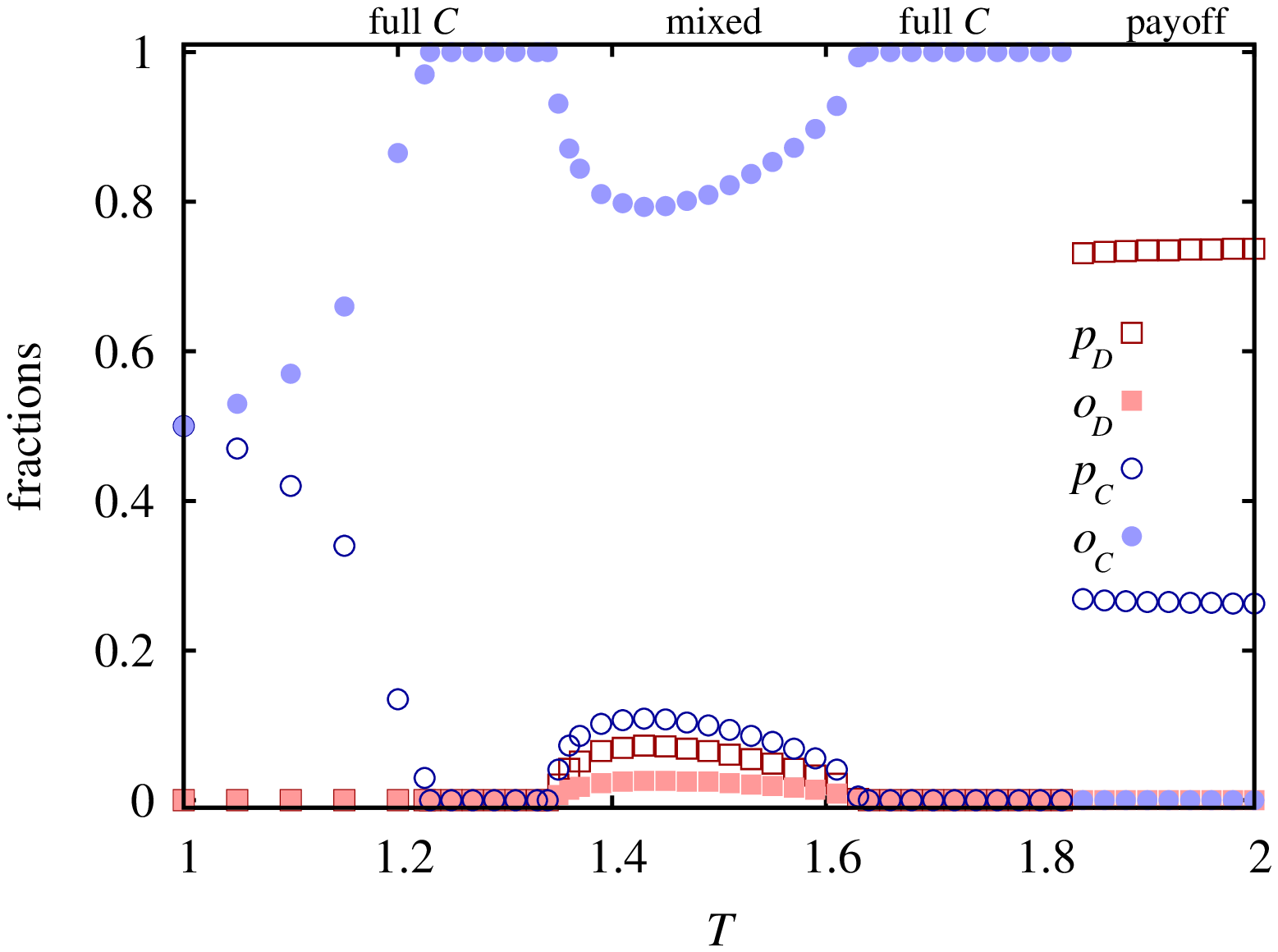}
\includegraphics[width=0.45\linewidth]{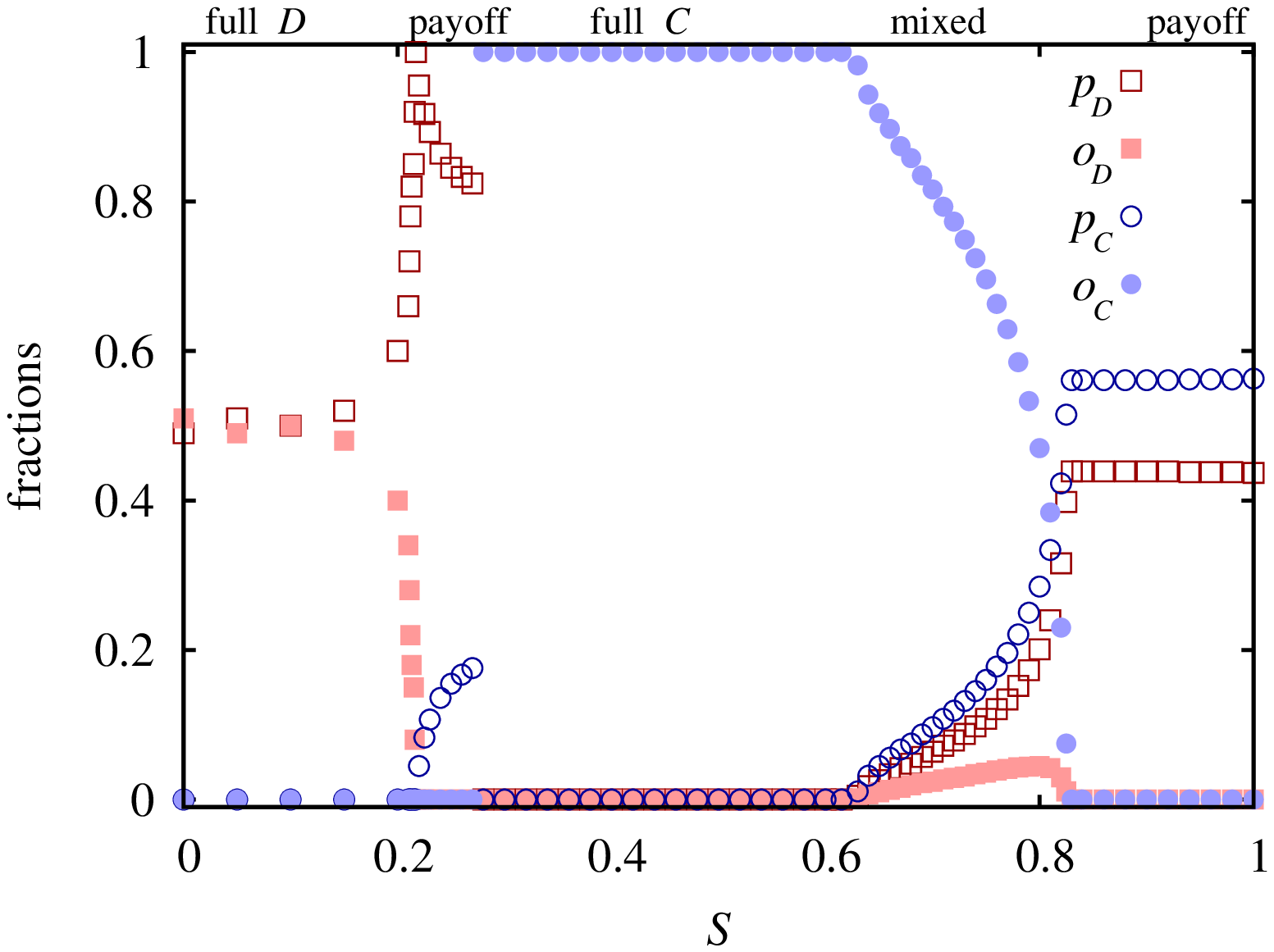}
\caption{Representative cross sections of the snowdrift phase diagram shown in Fig.~\ref{phd_sd}, as obtained for $S=0.7$ (left panel) and for $T=1.5$ (right panel). Depicted are stationary fractions of four states dependent on temptation $T$ (left) and sucker's payoff $S$ (right). Stable solutions are denoted along the top axis. Further description of emerging phase transitions can be found in the main text.}
\label{cross}
\end{figure}

For a more quantitative description we present two characteristic cross sections of the diagram in Fig.~\ref{cross}. In the left panel we fixed $S=0.7$ and increased the temptation value gradually. At small $T$ values both types of defectors die out soon and only $p_C$ and $o_C$ players survive. Here the system always evolves into a homogeneous state in the long run, but the likelihood to reach a homogeneous $p_C$ or a homogeneous $o_C$ state depends on the initial fractions of these players when neutral coarsening starts \cite{cox_ap86}. Consequently, the plotted fractions of $o_C$ and $p_C$ players in the early full $C$ region show simply the probability to reach the related homogeneous $o_C$ and $p_C$ states. As we increase $T$ the four-state mixed solution emerges, followed by a full cooperator state again, and finally the coexistence of $p_C$ and $p_D$ players becomes stable at high $T$ values. While the emergence and decline of the mixed solution happens via a continuous phase transition the change from a full $C$ state to a payoff-driven solution is always discontinuous.

In the right panel of Fig.~\ref{cross} we fixed $T=1.5$ and increased the sucker's payoff gradually. At small $S$ values cooperator players die out very soon and only $p_D$ and $o_D$ players survive. Similarly to the full $C$ state here the system always evolves into a homogeneous $p_D$ or a homogeneous $o_D$ state via a slow coarsening. Accordingly, the plotted fractions of $p_D$ and $o_D$ denote only the probability to reach the related homogeneous states. By increasing $S$ the stable coexistence of payoff-driven players emerges, which is replaced by the full dominance of $o_C$ players. This transition is always discontinuous. Higher $S$ offers a chance for all kind of players to survive that is followed by the dominance of payoff-driven learning protocol at very high $S$ values.

\begin{figure}
\centering
\includegraphics[width=0.32\linewidth]{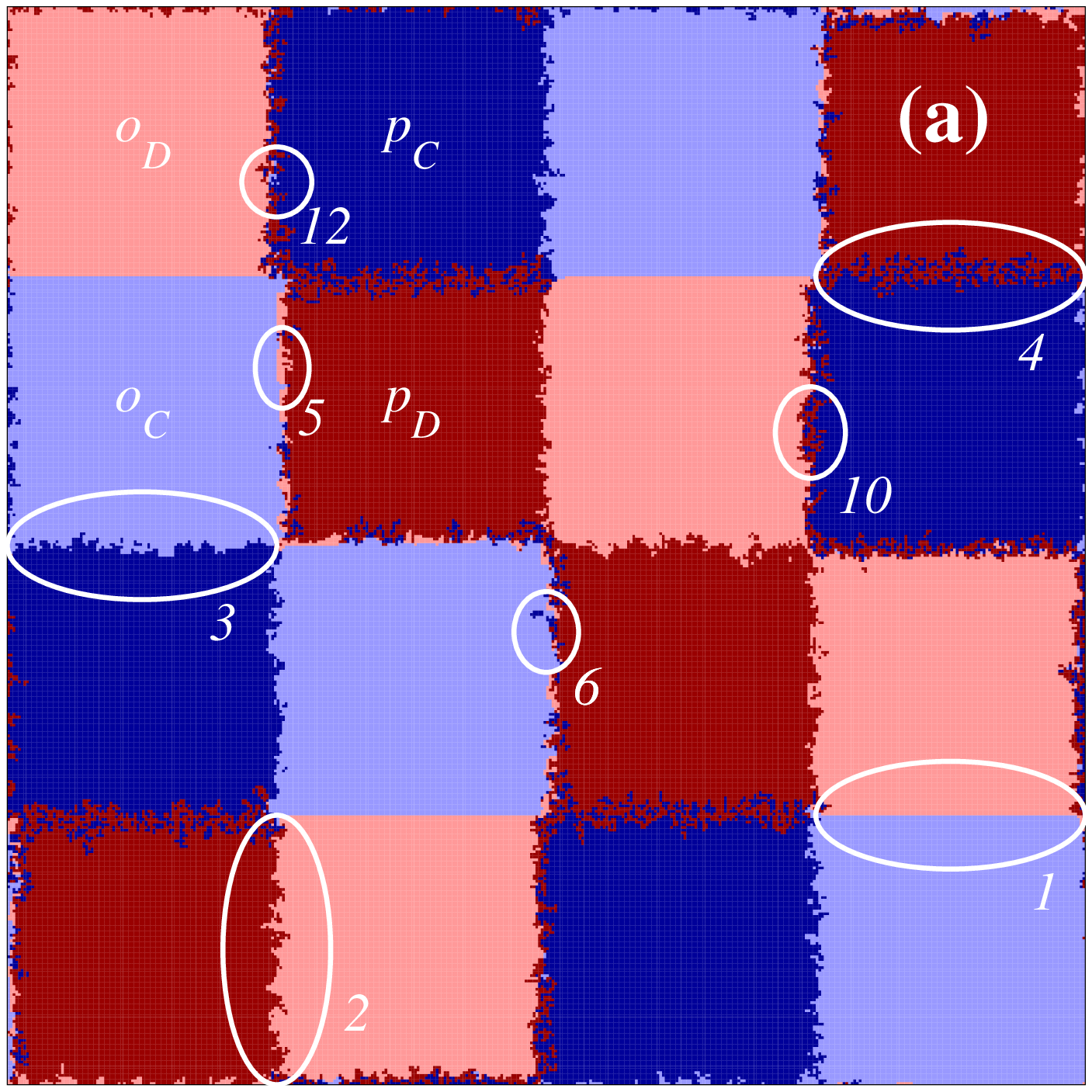}
\includegraphics[width=0.32\linewidth]{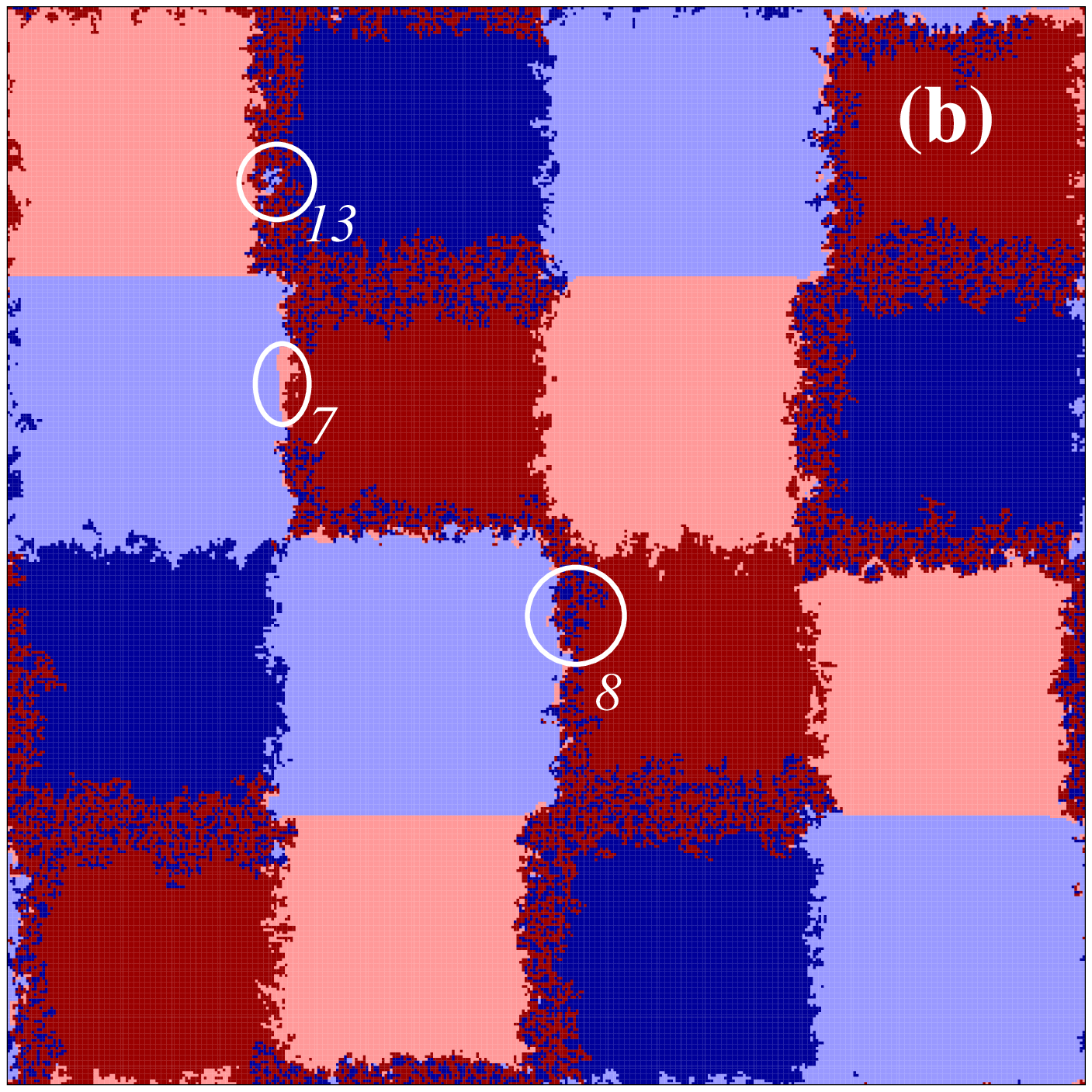}
\includegraphics[width=0.32\linewidth]{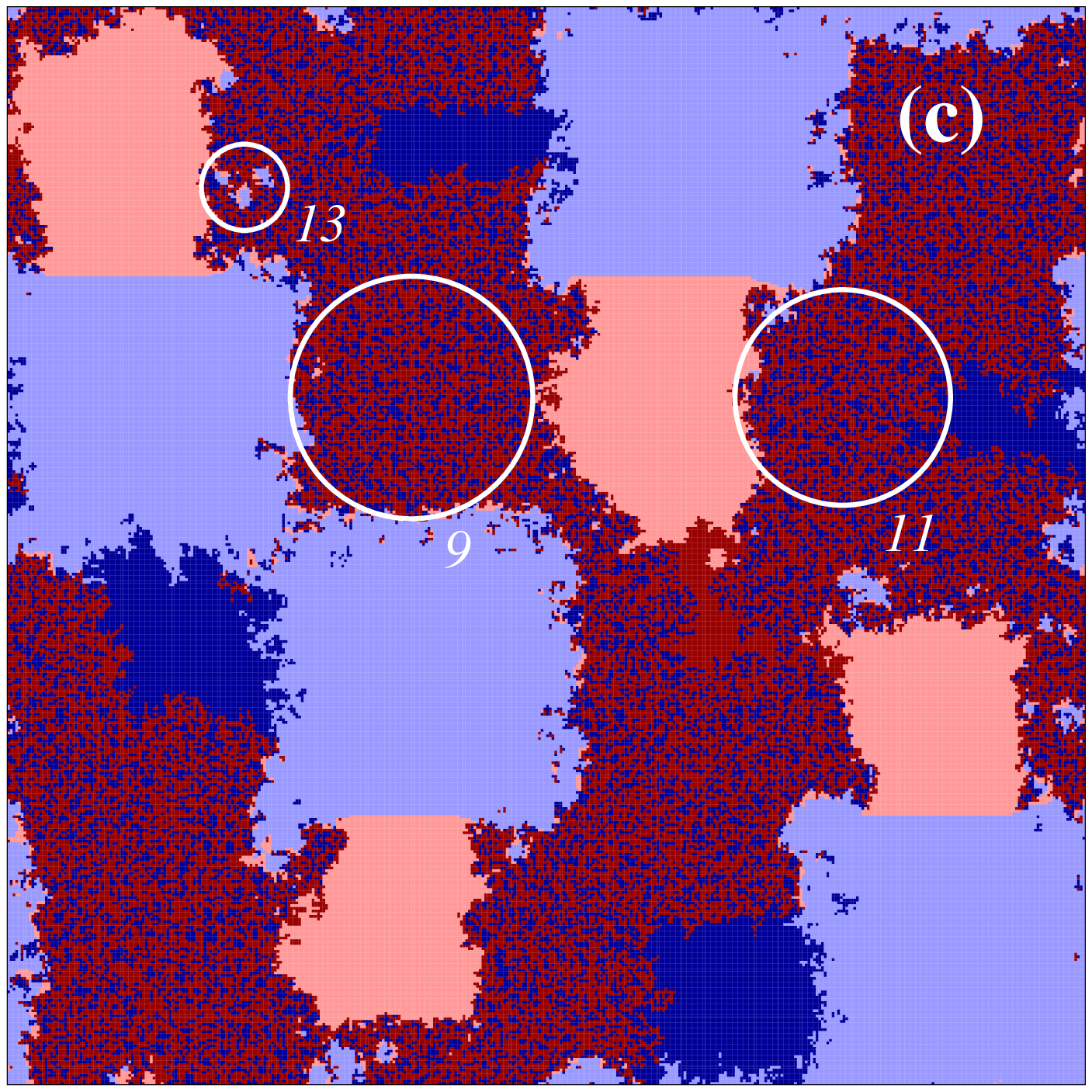}\\
\includegraphics[width=0.32\linewidth]{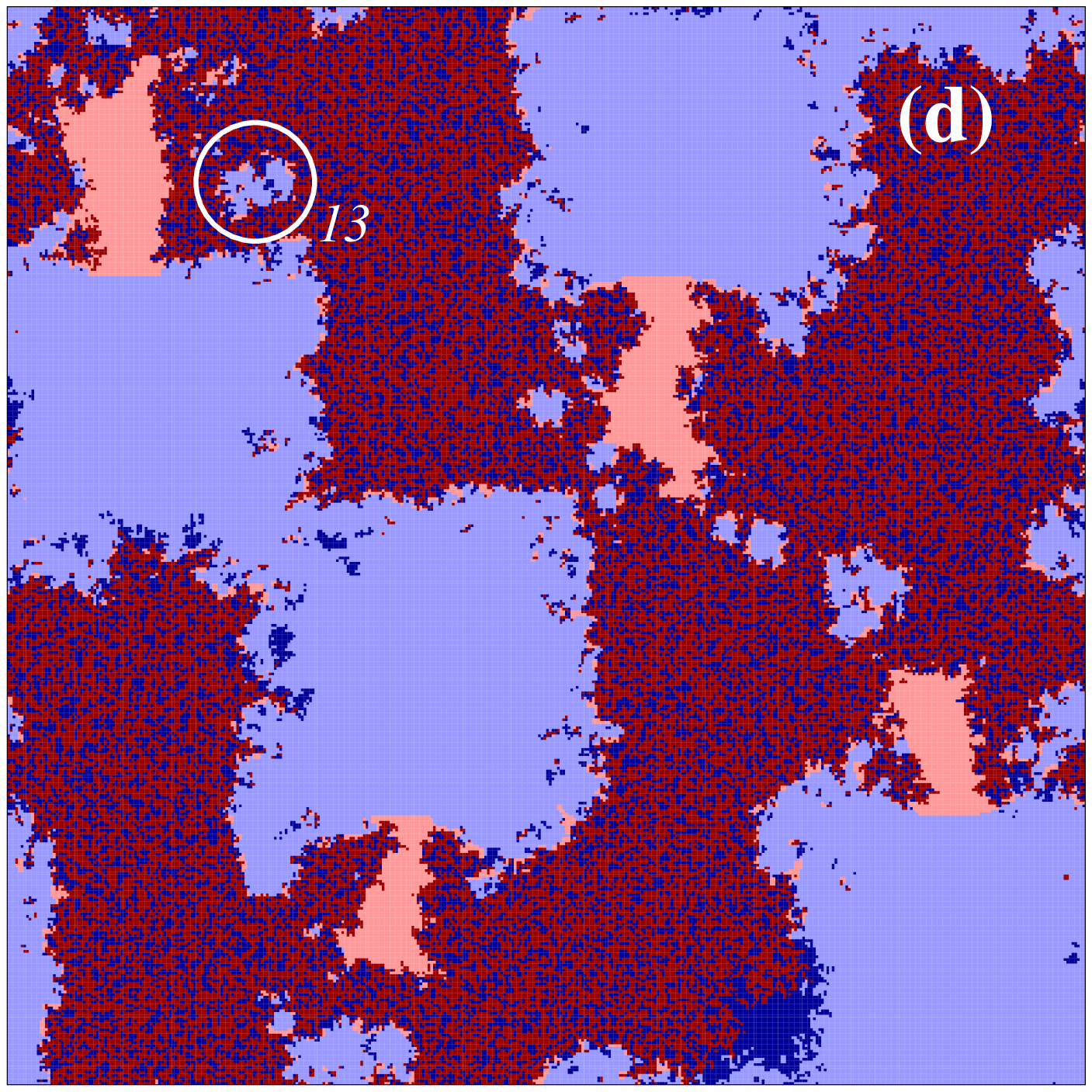}
\includegraphics[width=0.32\linewidth]{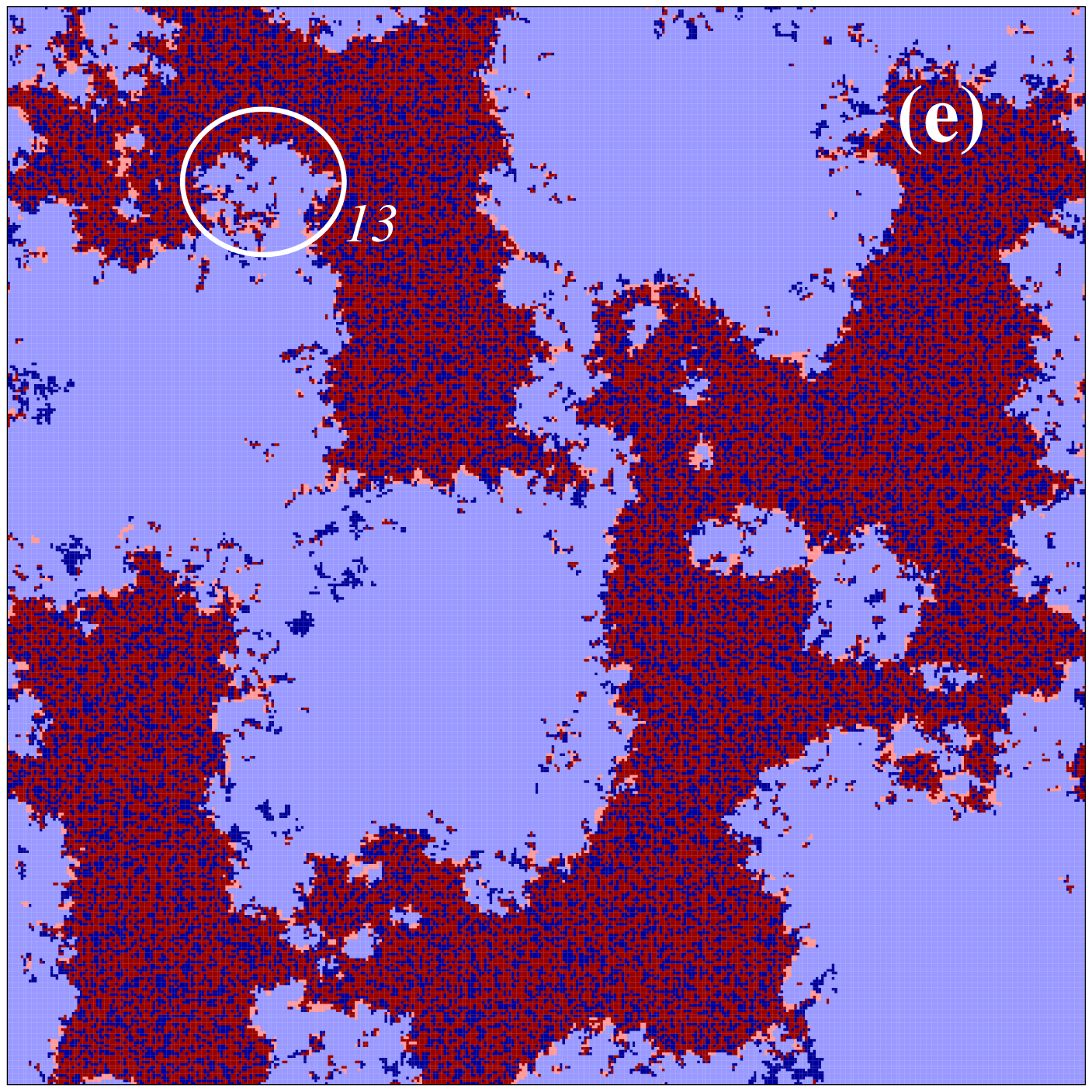}
\includegraphics[width=0.32\linewidth]{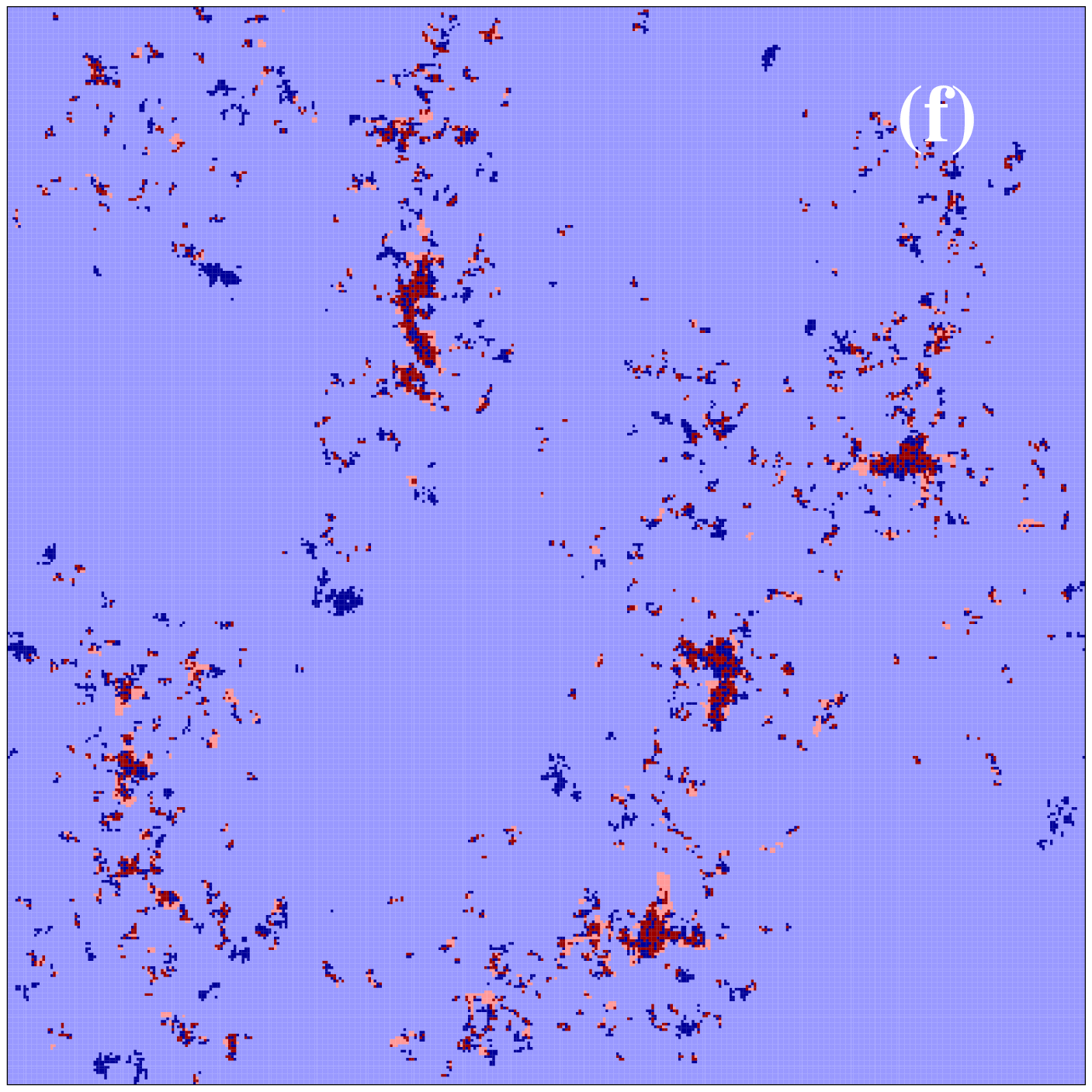}
\caption{Spatial evolution of the four competing states in a $400 \times 400$ system at $T=1.4$ and $S=0.4$ where the simulation is launched from a prepared patch-like initial state where all states are distributed in homogeneous domains (not shown). Here dark (light) blue denotes payoff-driven (conformity-driven) cooperators, while dark (light) red marks payoff-driven (conformity-driven) defectors, as it is marked by white labels in panel~(a). This panel represents the early stage of evolutionary process after 20 $MCS$s. Further stages of the evolutionary process are shown at 50, 150, 250, 350, and 800 $MCS$s. Finally the system evolves into a homogeneous $o_C$ state (not shown). The details of representative pattern formation processes are described in the main text. }
\label{snapshots_sd1}
\end{figure}

To get a deeper insight about the pattern formations which resulted in the above described evolutionary outcomes we present a representative evolution of spatially distributed players in Fig.~\ref{snapshots_sd1}. Here we do not use the traditional random initial distribution of available microscopic states, but apply a specially prepared patch-like state where all kind of interfaces between competing players can be found. In this way we can monitor all emerging pattern formations simultaneously via a single run. The starting state, where all borders are flat between homogeneous patches, is not shown, but panel~(a) of Fig.~\ref{snapshots_sd1} shows an early stage of evolutionary process. This panel illustrates nicely that the flat border between $o_D$ and $o_C$ remains practically frozen, as it is highlighted by a white ellipse and marked by $``1"$, because both strategies support their akin players at the front by ensuring the majority of similar strategies around them. Similarly, the border between $o_D$ and $p_D$ domains, denoted by ellipse $``2"$ and the border between $o_C$ and $p_C$ players (ellipse $``3"$) do not really propagate, but just fluctuate due to a neutral voter-model like slow coarsening. Alternatively, the interface between $p_C$ and $p_D$ domains, marked by ellipse $``4"$, starts diffusing intensively yielding a stable coexistence between these players. Note that in a two-player subsystem, which is identical to the traditional payoff-driven uniform system, these players would coexist at these $T-S$ values.

Interestingly, new states emerge at the front between $p_D$ and $o_C$ domain, which were not initially present. On one hand $p_D$ adopts the conformity attitude from $o_C$ because the latter reaches higher payoff and becomes $o_D$, shown by ellipse $``5"$. On the other hand, $p_D$ may adopt the strategy of $o_C$ and becomes $p_C$, as it is illustrated by ellipse $``6"$. However, due to the neutral relation between defector states the emerging $o_D$ cannot spread in the bulk of $p_D$, but stick at the original border of $o_C$ and $p_D$ , as it is shown by ellipse $``7"$ in panel~(b) of Fig.~\ref{snapshots_sd1}. Noteworthy,
$p_D$ players can enter into the bulk of $p_C$ players and build up a stable coexistence, as shown by ellipse $``8"$ in panel~(b). In this way the original border between $p_D$ and $o_C$ serves as a source of emerging $p_C+p_D$ solution which invades the pure $p_D$ domain - this is nicely demonstrated by ellipse $``9"$ in panel~(c).

Similarly to the above discussed case, the missing two microscopic states can also emerge at the original border of $p_C$ and $o_D$ domains. First, $p_C$ cannot reach really high payoff at the border, but $o_D$ is less vulnerable no matter her low payoff because she learns in a conformity-driven way. As a result, $p_C$ may change to $p_D$ state by learning the strategy of $o_D$, as highlighted by ellipse $``10"$ in panel~(a). After, as described previously, the emerging $p_D$ state can form a viable coexistence with $p_C$ players, hence the $p_C+p_D$ solution starts forming again. This is illustrated by ellipse $``11"$ in panel (c). Speaking about the fourth state, $o_C$ can also emerge at the original border of $o_D$ and $p_C$ because the fluctuating and highly disordered border may allow $o_D$ players to follow the strategy of $p_C$ and become $o_C$. This process is marked by ellipse $``12"$ in panel~(a). While the latter state remains viable, but its spread into the bulk of the $p_D+p_C$ solution is much slower than the propagation of $p_C+p_D$ solution in the original dark blue domain. (The slow growth of $o_C$ phase in $p_C+p_D$ solution is marked by ellipse $``13"$ in panels~(b,c,d) and (e)).

Summing up our observations, in the snowdrift quadrant practically the $p_C+p_D$ solution fights against the $o_C$ state. This is nicely demonstrated in panel~(e) of Fig.~\ref{snapshots_sd1}, where both solutions have already won their local battles and only these solutions survive to fight further for the final triumph. Since $o_C$ players follow conformity-biased learning protocol the enforced homogeneity of cooperation provides a highly competitive payoff that could be attractive for payoff-driven competitors. As a result, they become conformist easily. Contrarily, homogeneous $o_D$ domain cannot offer high payoff for payoff-driven players and the latter group stays at their original learning protocol. In this way the coevolution of payoff-driven and conformity-driven protocols can break the original symmetry of conformity-driven strategies, where both uniform destinations are possible, and pave the way for a full cooperator state that would be unreachable for the traditional model of uniform payoff-driven dynamics at such a high temptation value.
(The final evolutionary outcome, which is a homogeneous $o_C$ state, is not shown in our figure, but can be monitored in the animation provided in Ref.~\cite{sd}.)

\begin{figure}
\centering
\includegraphics[width=0.32\linewidth]{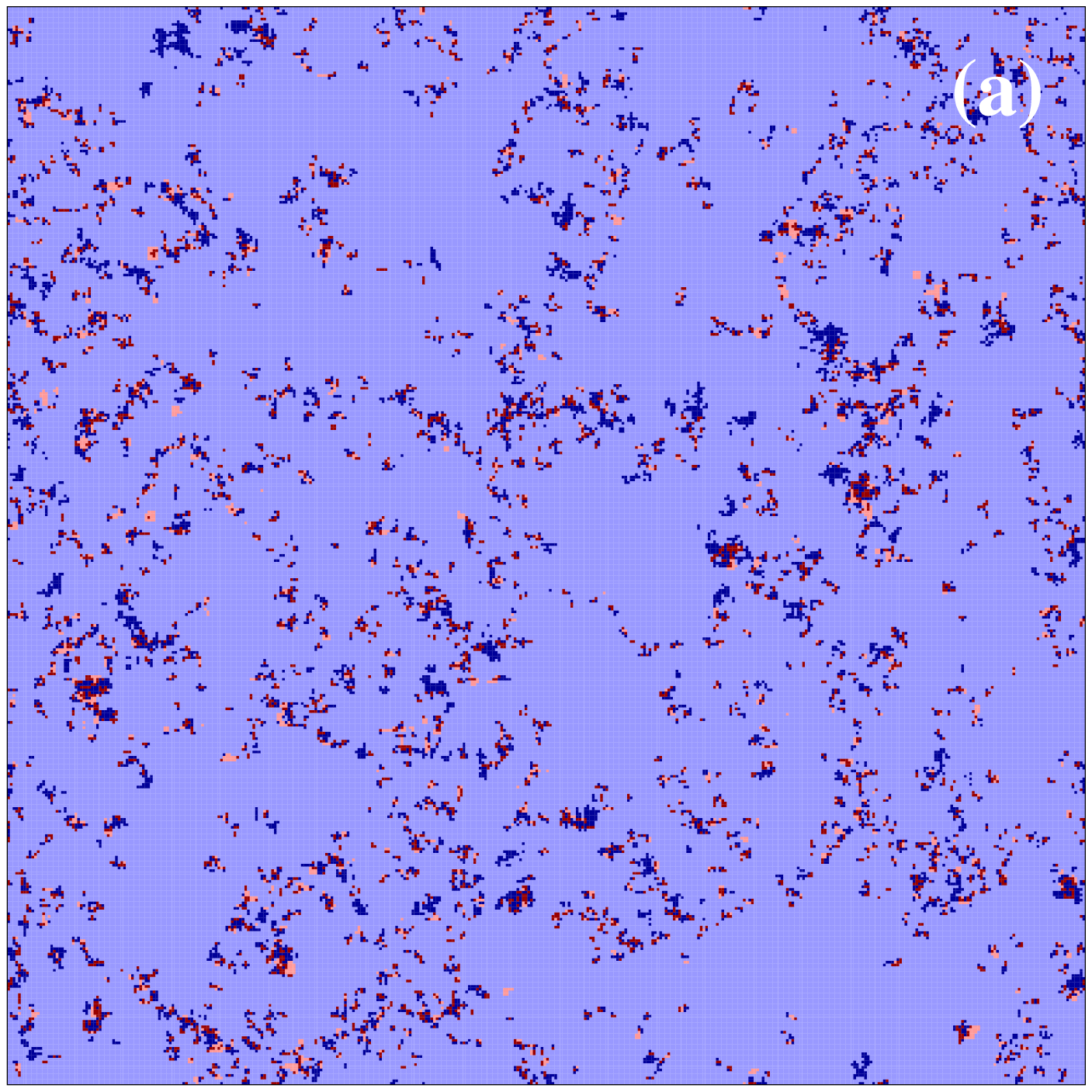}
\includegraphics[width=0.32\linewidth]{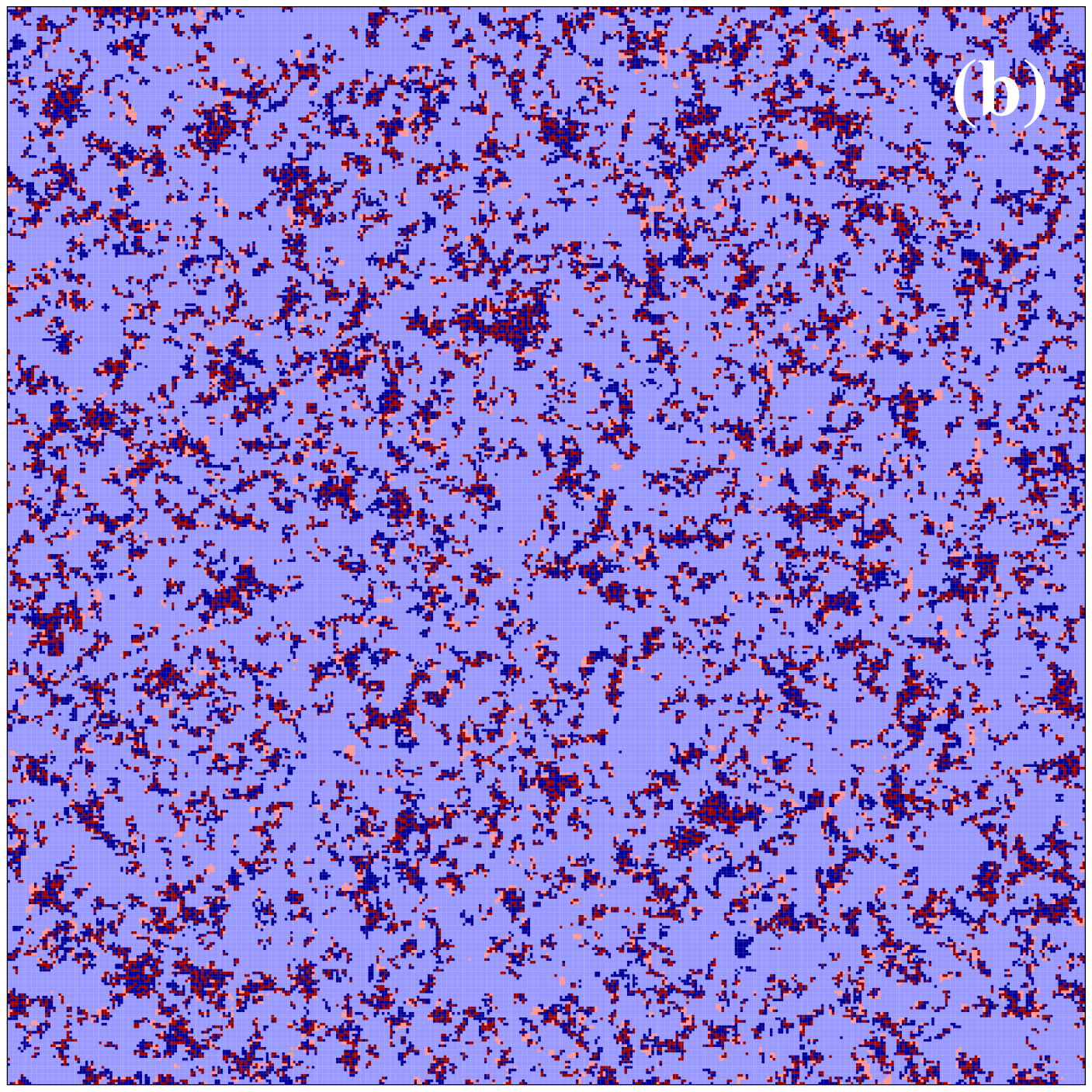}
\includegraphics[width=0.32\linewidth]{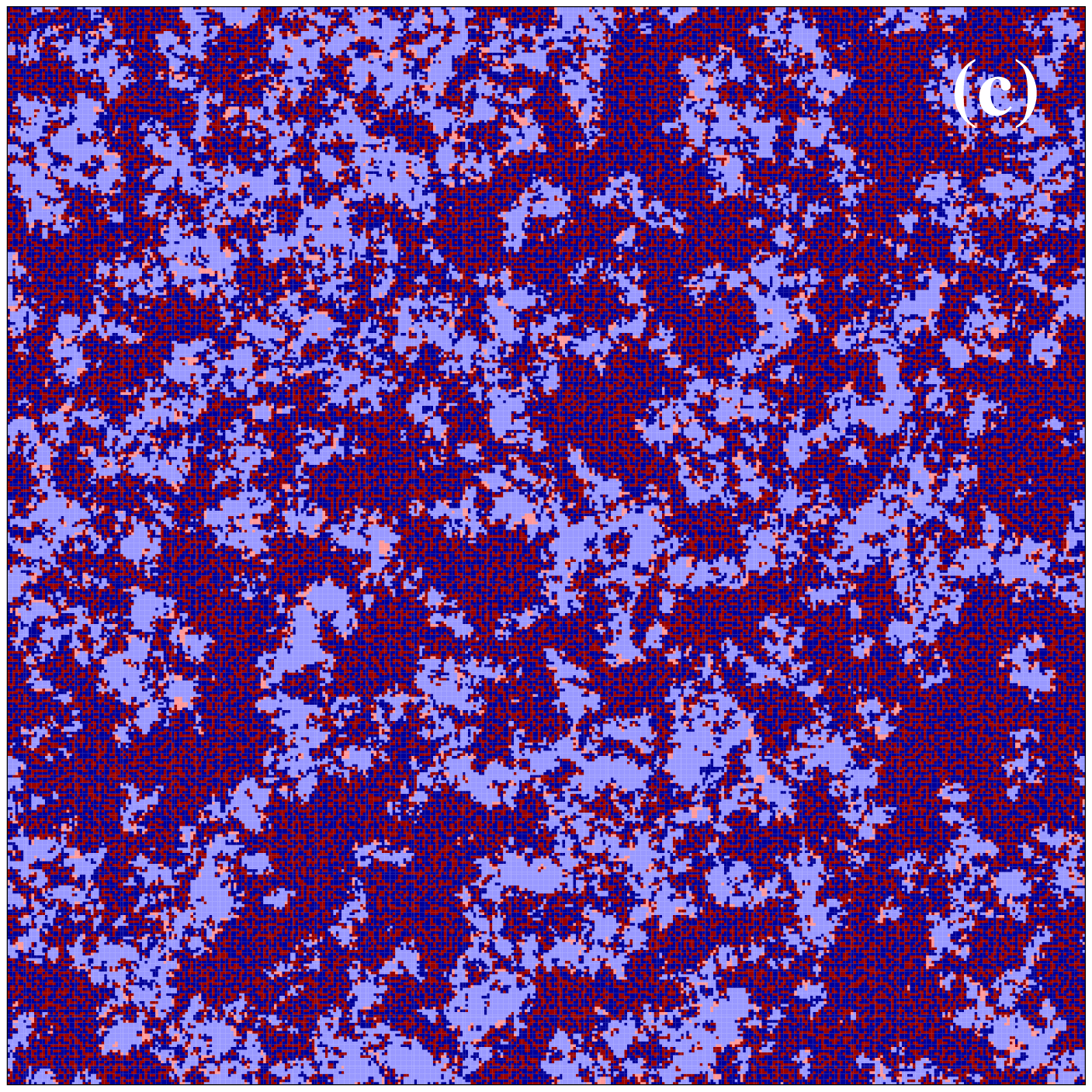}
\caption{Stationary distribution of different microscopic states obtained at $T=1.5$ for $S=0.65, 0.75$ and 0.82 from left to right. The color coding of different microscopic states are identical to those applied in Fig.~\ref{snapshots_sd1}. Namely, dark (light) blue denotes payoff-driven (conformity-driven) cooperators, while dark (light) red marks payoff-driven (conformity-driven) defectors. The $p_C+p_D$ domains become more and more dominant as we increase $S$ and above a threshold value this solution prevails in the whole space. Similar behavior may be detected at fixed $S$ as we gradually increase the value of $T$. Linear system size is $L=400$.}
\label{sd2}
\end{figure}

Our argument that makes clear the broader appearance of full cooperator state can also explain why payoff-driven solution becomes dominant for high $S$ and/or high $T$ values. In the latter cases the collective payoff of role-separating $p_C+p_D$ coexistence can reach the competitive payoff of $o_C$ domains. Consequently payoff-driven players can survive and they become dominant as we increase $T$ or $S$ further. The latter is illustrated in Fig.~\ref{sd2}, where we plotted three representative snapshots of the stationary state for three different $S$ values where all microscopic states are present in the ``mixed" phase. These plots highlight that the $p_C+p_D$ solution gradually crowd out $o_C$ state as we increase $S$ and becomes exclusive above a critical value of $S$. Notably, as we already discussed above, the interface between $o_C$ and $p_C+p_D$ domains offers a chance for $o_D$ state to emerge which explains the presence of all available microscopic states in the ``mixed" phase. Last, we note that similar behavior can also be reached at fixed $S$ by increasing $T$ only, because the competitive payoff for $p_C+p_D$ solution can be ensured in the latter way, too.

Turning to the prisoner's dilemma quadrant of $T-S$ plane, the application of our coevolutionary model cannot yield notably different results from the classical case when players apply payoff-driven learning method uniformly \cite{szabo_pr07}. More precisely, both $o_C$ and $p_C$ players die out soon after we launch the evolution from a random initial state and only defectors survive. 
In other words, deep in the prisoner's dilemma quadrant when $T>1$ and $S<0$ the coevolution of different learning methods cannot yield additional help for cooperators to survive. 

\begin{figure}
\centering
\includegraphics[width=0.7\linewidth]{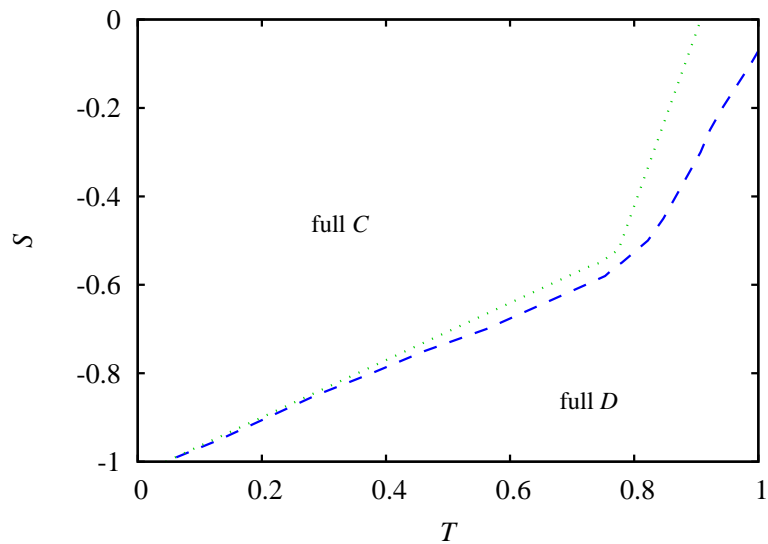}
\caption{Full $T-S$ phase diagram of stag-hunt game where payoff-driven and conformity-driven strategy learning collaborate to beat defectors. Dashed blue line shows the border of full cooperator and full defector phases. For comparison dotted green line shows the same border for the classic model when only payoff-driven strategy learning is used.}
\label{phd_sh}
\end{figure}

The evolutionary outcome, however, is more interesting in the stag-hunt quadrant where $T<1$ and $S<0$. Here in the classic case either defectors or cooperators prevail depending on the actual $T-S$ values \cite{szabo_jtb12}. This evolutionary outcome remains valid for the present coevolutionary model, but now cooperators can dominate larger parameter areas from defectors. This is illustrated in Fig.~\ref{phd_sh} where the full cooperator and full defector states are separated by a discontinuous phase transition. For comparison we have also plotted the same border line in the case when only payoff-driven strategy learning is used by the players. The comparison suggests that in the coevolutionary case there is a significant area of $T-S$ plane where the direction of the evolutionary process can be reversed and a full cooperator state is reached instead of full defection destination.

\begin{figure}
\centering
\includegraphics[width=0.32\linewidth]{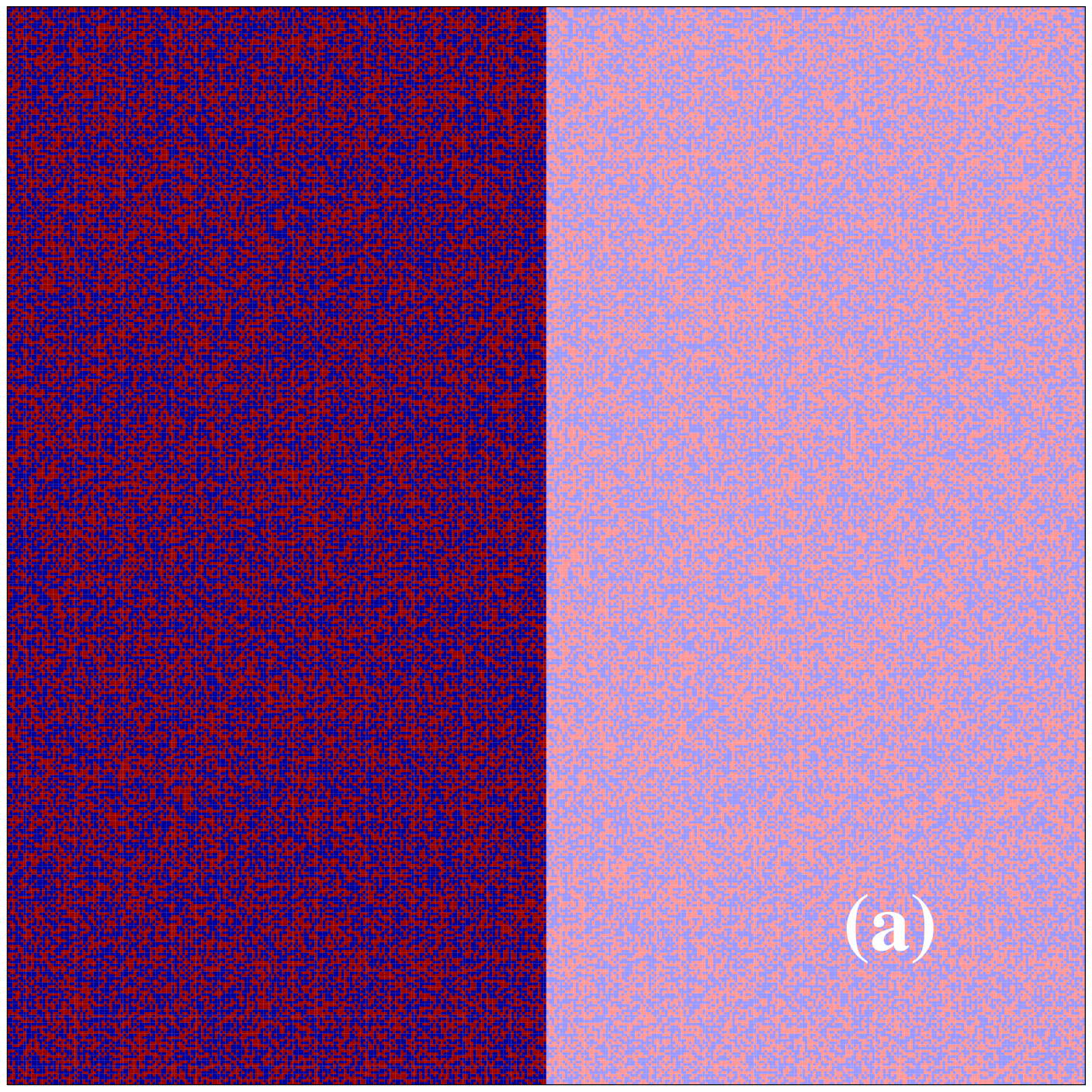}
\includegraphics[width=0.32\linewidth]{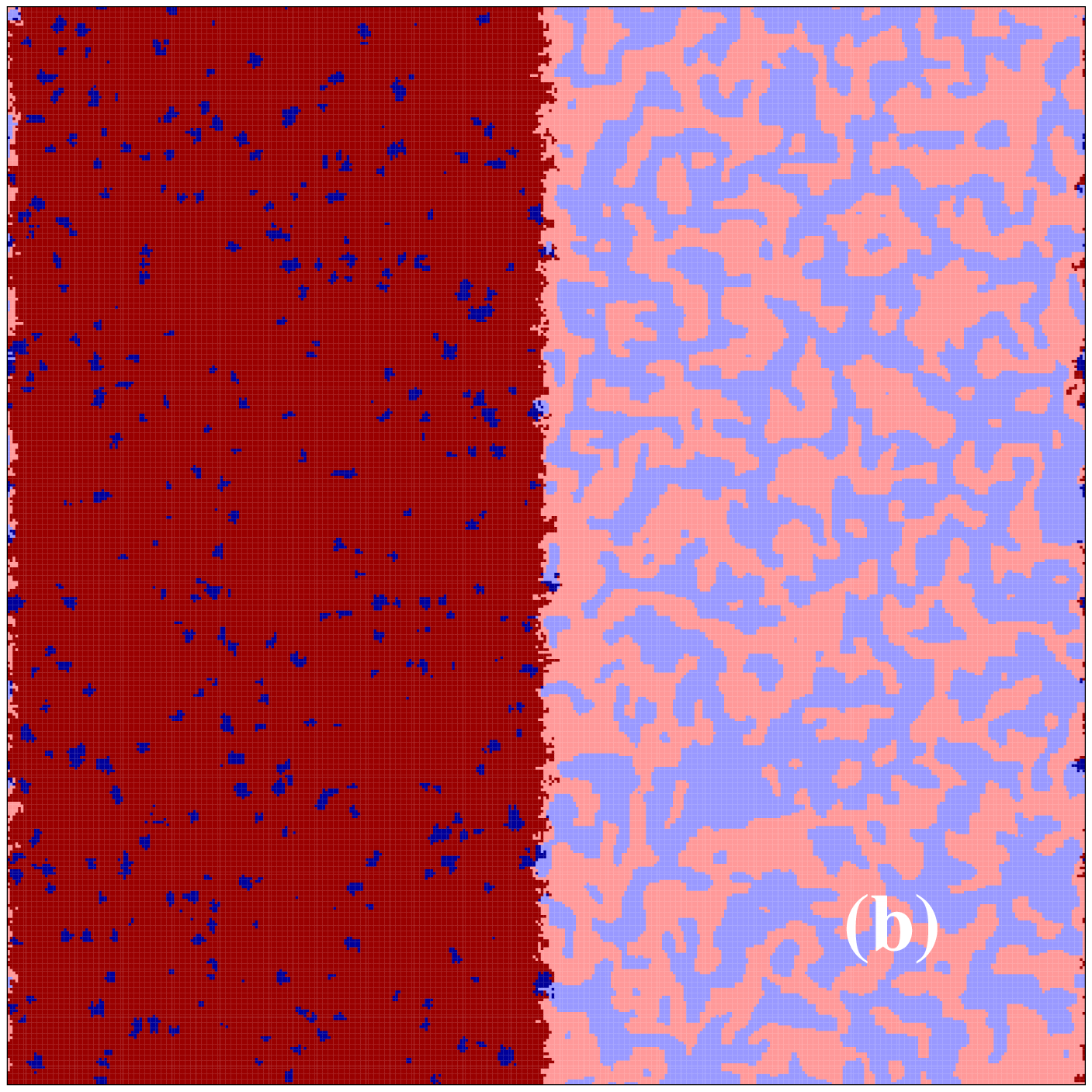}
\includegraphics[width=0.32\linewidth]{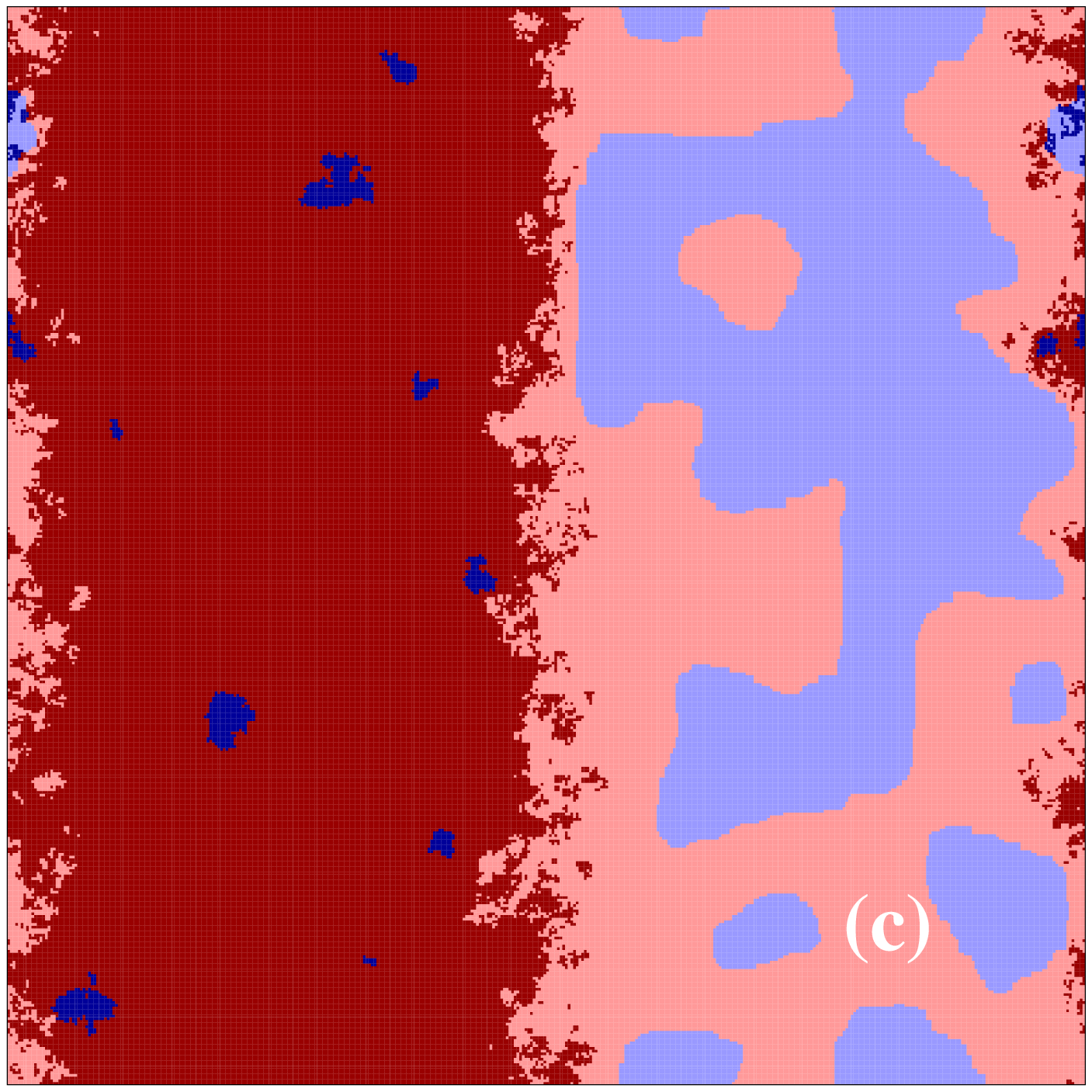}\\
\includegraphics[width=0.32\linewidth]{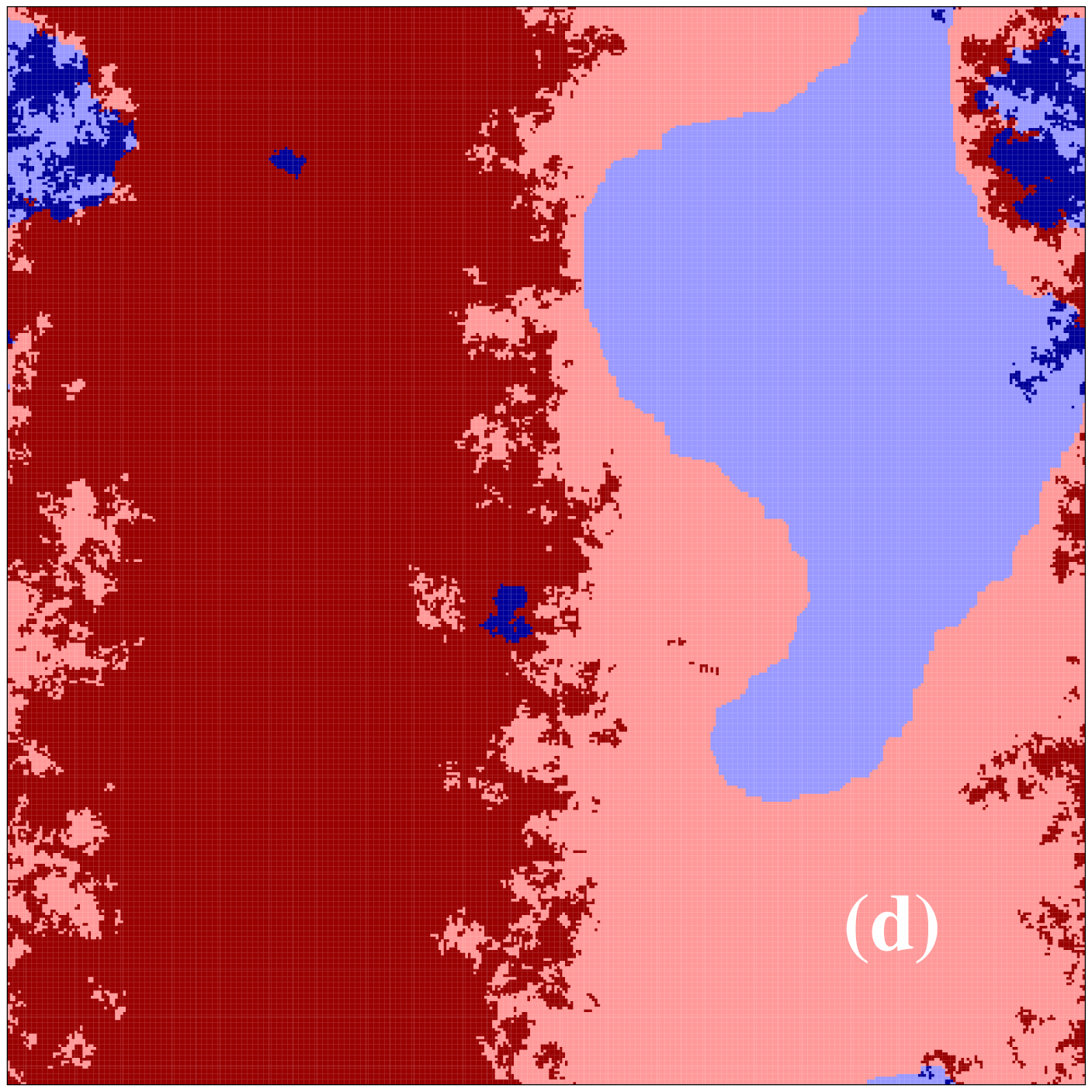}
\includegraphics[width=0.32\linewidth]{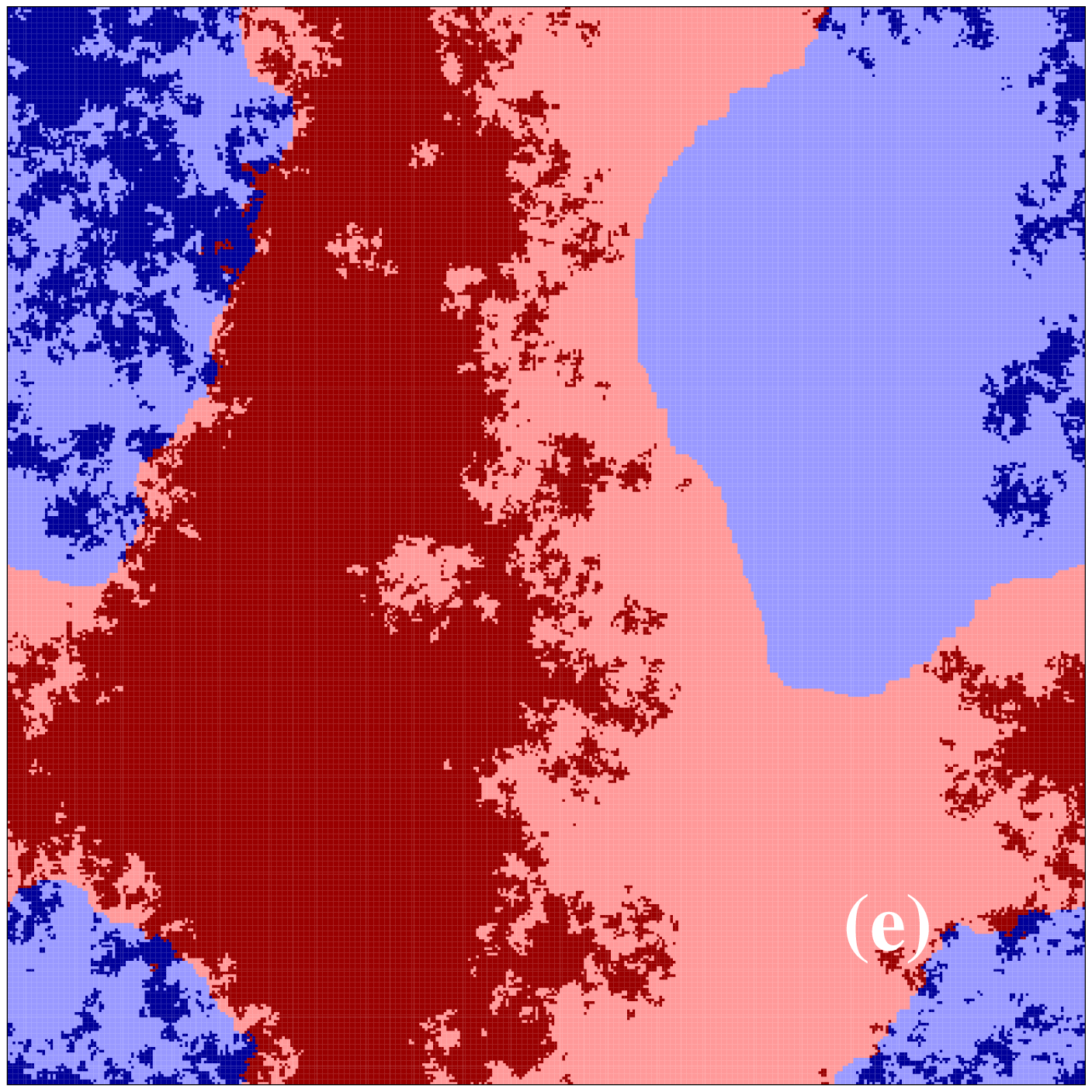}
\includegraphics[width=0.32\linewidth]{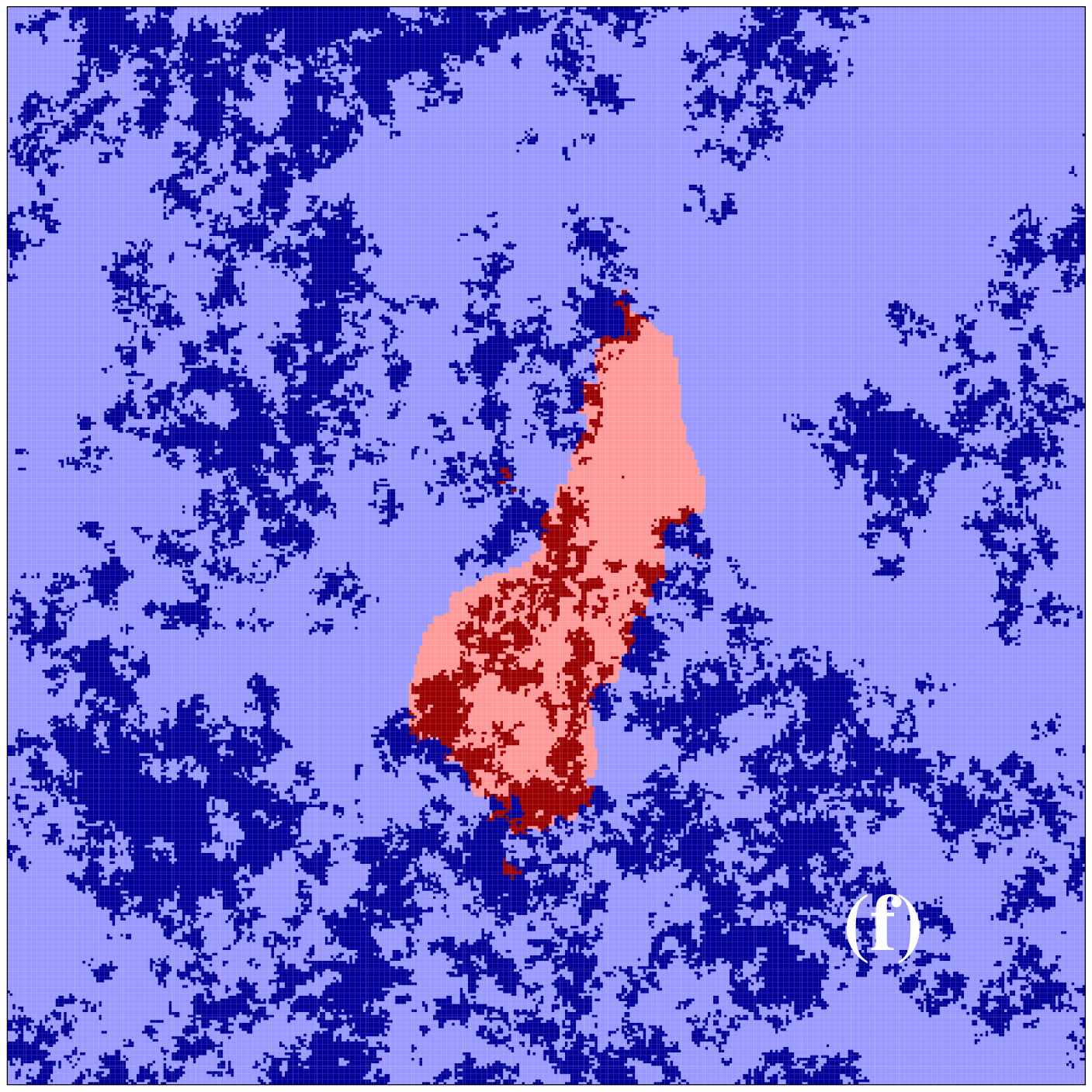}
\caption{Spatial evolution starting from a random initial state where players are separated according to their learning methods. The color code of different states are identical to the case we used in Fig.~\ref{snapshots_sd1}. In panel~(a) the $400 \times 400$ system is divided into two halves, where left side is occupied by payoff-driven players while the right side is occupied by conformity-driven players. First, red color defectors seem to dominate, but after the winning $o_C+p_C$ (mixture of light and dark blue colors) alliance emerges in panel~(d) and the direction of evolution is reversed. Finally the system terminates onto a full $C$ state (not shown) that is unreachable for classic model at the applied $T=0.89$, and $S=-0.3$ parameter pair. The different stages of the evolutionary process are shown at 20, 600, 2100, 5500, and 11500 $MCS$s, respectively. Further details of emerging pattern formation are given in the main text.}
\label{snapshots_sh}
\end{figure}

To understand the origin of this remarkably different outcome we present a series of snapshots of evolution starting from a prepared initial state at $T=0.89$ and $S=-0.3$. In the initial state, shown in panel~(a) of Fig.~\ref{snapshots_sh}, players using payoff-driven learning method are distributed randomly in the left side of the space while players using conformity-driven strategy learning method are distributed on the right side. The early stage after the evolution is launched can be seen in panel~(b). It suggests that $p_D$ players can easily beat $p_C$ players and the final outcome of this subsystem would be a full defector state. (The last surviving dark blue domain of $p_C$ players in the sea of dark red $p_D$ players goes extinct in panel~(d).) Notably, the evolution in the right side is different because the strategy-neutral conformity-driven microscopic rule would allow both destinations of uniform states. In this sub-system the curvature-driven coarsening determines how domains become larger and larger by eliminating the peaks and bulges from the interfaces separating homogeneous $o_C$ and $o_D$ domains. Consequently, small islands are always shrinking in the bulk of a larger domain. Only straight frontiers may be stable temporarily but their edges, where the curvature is relevant, are also unstable. The vicinity of payoff-driven players, however, breaks the original symmetry of $o_C$ and $o_D$ states. Since payoff-driven players are mostly defectors, neighboring conformity-driven players will also prefer defector strategy. As a result, light-blue $o_C$ domains would gradually disappear and defection seems to be a victor in both sub-systems. 

Interestingly, however, the mixture of $o_C$ and $p_C$ players (light and dark blue) can form an effective alliance against defectors. While $o_C$, surrounded by other cooperators, is less vulnerable against $p_D$ players, $p_C$ can also utilize that $o_D$ players need a regular interface for invasion. In the absence of it the latter becomes also susceptible to the vicinity of $p_C$ players. These two effects altogether establish an effective alliance of different types of cooperator strategies, who can gradually prevail in the whole system (the final destination to a full cooperator state is not shown in Fig.~\ref{snapshots_sh}, but can be monitored in the animation provided in Ref.~\cite{sh}). It is worth noting that the mixture of $o_D$ and $p_D$ players cannot form similar effective alliance because $p_D$ state cannot utilize the vicinity of $o_D$ players.

\section{Discussion}

In this work we have studied the coevolution of competing strategies and their learning methods. Beside the broadly applied payoff-driven imitation dynamics we have also assumed players to use conformity-driven learning and allowed them to change their learning protocols in time. The proper coevolution revealed significantly different behaviors from those cases when we just apply the mentioned learning methods simultaneously in a static way via an external control parameter. In the latter cases players can only vary their strategies during the evolution \cite{yang_hx_csf17,liu_rr_epl15,du_jm_fp18,takesue_epl18}. In other words, the significance of present study is to reveal the qualitatively different pattern formation mechanisms that we can only observe in a coevolutionary framework.

We have shown that there are parameter regions, like strong snowdrift game situations at high $T$ or high $S$ values where the competition of different learning methods results in the unambiguous victor of payoff-driven strategy learning. Here the role separating cooperator-defector pairs provide so high collective payoff value that cannot be beaten by a homogeneous domain which would be a consequence of a conformity-driven learning method. Nevertheless, for less sharp snowdrift game regions at smaller $T$ and $S$ values the coevolution of different learning methods is useful to reach a full cooperative state that would not be reachable otherwise. In the latter case the homogeneous cooperator domains can invade the whole space by enjoying the advantage of conformity-driven learning method.

Interestingly, in the stag-hunt game region the simultaneous presence of different learning methods reveals a novel way of collaboration that cannot be observed otherwise. Here conformity-driven cooperators can resist the invasion of payoff-driven defectors and neighboring payoff-driven cooperators can attack conformity-driven defectors successfully. The expected symmetry is broken for defectors because $o_D$ and $p_D$ states cannot form similarly efficient alliance. In this way the active partnership of different types of cooperator players allows them to extend the full cooperator state to those parameter values which belonged to the sovereignty of defectors in the classic payoff-driven model.

The diversity of players has already been proved to be useful to maintain cooperation in harsh environment where defection would prevail in a homogeneous population \cite{santos_prl05,szolnoki_epl07,santos_n08,perc_pre08,fu_pre08,chen_yz_pre09,perc_njp11,javarone_jstat16,liu_ph_pa17b,yang_hx_jstat17,huang_cs_epl17}. Our present observations underline that this concept can be extended more generally in a coevolutionary framework \cite{perc_bs10,stivala_pre16,richter_bs17,zhang_w_pa17,wu_t_pcbi17} where the evolution either select one of the learning methods to prevail or allows coexistense by offering new solutions to emerge. Hopefully our extention can be useful for other kind of microscopic rules, including win-stay lose-shift, myopic, other-regarding preference, Pavlov-rule, or in general for those rules which use a sort of aspiration level for personal decision making \cite{posch_jtb99, posch_prslb99, chen_xj_pre09, fort_jsm05, taylor_c_tpb06, platkowski_pre09, perc_pone11, szabo_jtb12b, perc_srep15,wang_z_pa11, fu_mj_ijmpc18, wu_t_njp18, shen_rs18}.

\ack
This research was supported by the Hungarian National Research Fund (Grant K-120785) and  by the National Natural Science Foundation of China (Grants No. 61503062).

\providecommand{\newblock}{}

\end{document}